\documentclass[twocolumn,superscriptaddress,aps,preprintnumbers,amsmath,amssymb,prd,nofootinbib,pdftex]{revtex4-2}

\usepackage[pdftex]{graphicx}
\usepackage{hyperref}
\usepackage{xcolor}
\usepackage{here}
\usepackage{orcidlink}

\newcommand{\beq}{\begin{equation}}
\newcommand{\eeq}{\end{equation}}
\newcommand{\beqa}{\begin{eqnarray}}
\newcommand{\eeqa}{\end{eqnarray}}

\def\df{\mathrm{d}}

\newcommand{\fDM}{f_{\mathrm{DM}}}
\newcommand{\vvir}{v_{\mathrm{vir}}}

\newcommand{\e}{\mathrm{e}}
\newcommand{\iu}{\mathrm{i}}

\newcommand{\erf}{\mathrm{erf}}

\definecolor{forestgreen}{HTML}{228B22}
\usepackage{ulem}

\begin{document}

\title{Ultralight vector dark matter search using data from the KAGRA O3GK run
}
\author{A.~G.~Abac}
\affiliation{Max Planck Institute for Gravitational Physics (Albert Einstein Institute), D-14476 Potsdam, Germany}
\author{R.~Abbott}
\affiliation{LIGO Laboratory, California Institute of Technology, Pasadena, CA 91125, USA}
\author{H.~Abe}
\affiliation{Graduate School of Science, Tokyo Institute of Technology, 2-12-1 Ookayama, Meguro-ku, Tokyo 152-8551, Japan}
\author{I.~Abouelfettouh}
\affiliation{LIGO Hanford Observatory, Richland, WA 99352, USA}
\author{F.~Acernese}
\affiliation{Dipartimento di Farmacia, Universit\`a di Salerno, I-84084 Fisciano, Salerno, Italy}
\affiliation{INFN, Sezione di Napoli, I-80126 Napoli, Italy}
\author{K.~Ackley\,\orcidlink{0000-0002-8648-0767}}
\affiliation{University of Warwick, Coventry CV4 7AL, United Kingdom}
\author{C.~Adamcewicz\,\orcidlink{0000-0001-5525-6255}}
\affiliation{OzGrav, School of Physics \& Astronomy, Monash University, Clayton 3800, Victoria, Australia}
\author{S.~Adhicary}
\affiliation{The Pennsylvania State University, University Park, PA 16802, USA}
\author{N.~Adhikari\,\orcidlink{0000-0002-4559-8427}}
\affiliation{University of Wisconsin-Milwaukee, Milwaukee, WI 53201, USA}
\author{R.~X.~Adhikari\,\orcidlink{0000-0002-5731-5076}}
\affiliation{LIGO Laboratory, California Institute of Technology, Pasadena, CA 91125, USA}
% ...

\collaboration{The LIGO Scientific Collaboration, the Virgo Collaboration, and the KAGRA Collaboration}

\author{T.~Fujimori}
\affiliation{Department of Physics, Graduate School of Science, Osaka Metropolitan University, 3-3-138 Sugimoto-cho, Sumiyoshi-ku, Osaka City, Osaka 558-8585, Japan}

\author{H.~Fujimoto\,\orcidlink{0000-0002-0070-0678}}
\affiliation{Department of Physics, The University of Tokyo, 7-3-1 Hongo, Bunkyo-ku, Tokyo 113-0033, Japan}

\author{T.~Fujita\,\orcidlink{0000-0003-4722-7432}}
\affiliation{Waseda Institute for Advanced Study, Waseda University, 1-6-1 Nishi-Waseda, Shinjuku, Tokyo 169-8050, Japan}
\affiliation{University of Tokyo, Tokyo, 113-0033, Japan.}

\author{Y.~Manita\,\orcidlink{0000-0002-2673-8404}}
\affiliation{Department of Physics, Kyoto University, Kita-Shirakawa Oiwake-cho, Sakyou-ku, Kyoto City, Kyoto 606-8502, Japan}

\author{I.~Obata\,\orcidlink{0000-0001-9737-5631}}
\affiliation{Kavli Institute for the Physics and Mathematics of the Universe, The University of Tokyo, 5-1-5 Kashiwa-no-Ha, Kashiwa City, Chiba 277-8583, Japan}

\author{H.~Takidera}
\affiliation{Department of Physics, The University of Tokyo, 7-3-1 Hongo, Bunkyo-ku, Tokyo 113-0033, Japan}

\begin{abstract}
Among the various candidates for dark matter (DM), ultralight vector DM can be probed by laser interferometric gravitational wave detectors through the measurement of oscillating length changes in the arm cavities. 
In this context, KAGRA has a unique feature due to differing compositions of its mirrors, enhancing the signal of vector DM in the 
length change in the auxiliary channels. 
Here we present the result of a search for $U(1)_{B-L}$ gauge boson DM using the KAGRA data from auxiliary length channels during the first joint observation run together with GEO600. 
By applying our search pipeline, which takes into account the stochastic nature of ultralight DM, upper bounds on the coupling strength between the $U(1)_{B-L}$ gauge boson and ordinary matter are obtained for a range of DM masses. 
While our constraints are less stringent than those derived from previous experiments, this study demonstrates the applicability of our method to the lower-mass vector DM search, which is made difficult in this measurement by the short observation time compared to the auto-correlation time scale of DM.
\end{abstract}

\maketitle

\section{Introduction}
Recently, a number of novel dark matter (DM) searches using laser interferometric gravitational wave (GW) detectors have been proposed~\cite{Pierce:2018xmy,Nagano:2019rbw, Michimura:2020vxn,Miller:2020vsl,Nagano:2021kwx, Miller:2022wxu} and conducted~\cite{Guo:2019ker, Morisaki:2020gui, LIGOScientific:2021ffg, Vermeulen_2021, Miller:2023kkd}.
Because of their extremely high sensitivity to the differential length changes of their arms in the frequency range $\mathcal{O}(10 - 10^{3})$ Hz, they can probe the interaction between the detector and DMs, which have masses of $\mathcal{O}(10^{-14}- 10^{-11})$ eV/$c^2$ and therefore oscillate coherently within this frequency band.
Vector DM (or so-called dark photon DM) interacts with test masses of the interferometer, for example, via a coupling to the baryon ($B$) or baryon minus lepton ($B - L$) number current. 
Because of the non-relativistic dispersion (under the standard halo model~\cite{Bertone:2004pz,Evans:2018bqy}), the vector DM field exerted on the test masses behave as an oscillating dark electric force, inducing a measurable change in the differential length of the arm cavity. 
Vector DMs of this type, which we refer to as $U(1)_B$ and $U(1)_{B-L}$ gauge boson, were previously searched in Ref.~\cite{LIGOScientific:2021ffg} by using the latest observational data from the Advanced LIGO~\cite{LIGOScientific:2014pky, Tse:2019wcy, aLIGO:2020wna} and Virgo~\cite{VIRGO:2014yos, Virgo:2019juy} detectors. 
Remarkably, the constraint on the coupling strength of vector DM to baryons from GW interferometers surpasses those from existing experiments such as the E\"{o}t-Wash torsion balance~\cite{Schlamminger:2007ht, Wagner:2012ui} and MICROSCOPE~\cite{Touboul:2017grn, Berge:2017ovy, Fayet:2017pdp}, by orders of magnitude for certain frequency bands. 
Similarly, dilatonic DM, whose interaction alters the apparent electron mass or the fine structure constant, was probed with data from the GEO600 interferometer~\cite{Luck:2010rt, Dooley:2015fpa, Vermeulen_2021}. These searches highlight the potential of GW detectors as direct probes of ultralight DM.

A Japanese laser interferometric GW detector, KAGRA~\cite{Somiya:2011np, PhysRevD.88.043007, KAGRA:2020tym}, can also probe the vector DM interaction, but in a relatively unique way compared to the previous searches. 
For the Advanced LIGO and Virgo detectors, all the mirrors (including test masses) are made of the same material (fused silica), in other words, they have a common charge-to-mass ratio with respect to the dark electric field.
Because the spatial variation scale (so-called coherence length) of the ultralight DM field is $\mathcal{O}(10^{5} - 10^{8})$km in our target mass range, 
all mirrors of each detector, which are separated only a few kilo-meters, respond to the vector field nearly identically.
In contrast, KAGRA employs sapphire for cryogenic test masses and fused silica for room temperature auxiliary mirrors. 
Therefore, owing to the difference in the charge-to-mass ratios, those mirrors respond differently to the vector field, and the vector DM signal in the (differential) length change can be enhanced for auxiliary length monitors~\cite{Michimura:2020vxn,Michimura:2021hwr}.
Especially in the case of $U(1)_{B-L}$, where the difference in charge-mass ratio is relatively large, these channels are shown to have better sensitivity to vector DM than existing experiments in certain low frequency bands, when KAGRA reaches design sensitivity~\cite{Michimura:2020vxn}.

In this work, we conduct an ultralight vector DM search with KAGRA using the data of its auxiliary length change monitors during the first joint observation run together with GEO600 (O3GK)~\cite{KAGRA:2022twx}. Although the detector was in operation for two weeks during O3GK, the durations of the contiguous data segments were at most about 7 hours~\cite{KAGRA:2022fgc}. 
For our target mass range, this duration can be comparable to or smaller than the so-called coherence time, within which the amplitude and the phase of DM can be regarded as a constant. Since each such measurement is one realization of a random field, their stochasticity needs to be taken into account, especially when setting the upper bound on the coupling strength of the ultralight DM. 
Therefore, our pipeline is constructed based on a recent study~\cite{Nakatsuka:2022gaf}, which thoroughly investigates the stochastic nature of the ultralight vector DM. 
Using the detection statistic discussed in Ref.~\cite{Nakatsuka:2022gaf}, we numerically derive upper limits on the coupling strength incorporating the stochastic nature of the DM.
On the other hand, various noise lines~\cite{KAGRA:2022fgc} 
result in a false detection as a DM signal. In order to distinguish outliers, we have implemented veto procedures making use of expected features of ultralight DM signals.

The rest of the paper is organized as follows. We first introduce our model of ultralight vector DM and discuss its stochastic nature in Sec.~\ref{model}. After describing our search method in Sec.~\ref{method}, the results of our analysis using O3GK data are presented in Sec.~\ref{analysis}. Sec.~\ref{discussion} provides a discussion of these results and of the prospects for future searches.

\section{Dark matter search with KAGRA}\label{model}
\subsection{Vector Dark Matter Model}\label{vDMmodel}
In this work, we consider a ultralight vector dark matter field $A_{\mu}(t,x)$, which is regarded as a gauge boson of $U(1)_D$ gauge symmetry with $D$ being a label for a charge, such as $B$ and $B -L$. We assume it interacts with ordinary matter through the coupling to the $U(1)_D$ current $J^{\mu}_{D}$.
The Lagrangian density $\mathcal{L}$ is then given as
\beq
\mathcal{L} = -\frac{\varepsilon_0c^2}{4}F^{\mu\nu}F_{\mu\nu} + \frac{\varepsilon_0}{2}\left(\frac{m_Ac^2}{\hbar}\right)^2A^{\mu}A_{\mu}-\epsilon_{D}eJ^{\mu}_{D}A_{\mu},\label{eq:lagrangian}
\eeq
where $F_{\mu\nu} = \partial_{\mu}A_{\nu}-\partial_{\nu}A_{\mu}$ is the field strength, $c$ is the speed of light, $\hbar$ is the reduced Planck constant, $m_A$ is the mass of the vector field, $\varepsilon_0$ is the permittivity of vacuum and $\epsilon_D$ is the gauge coupling constant normalized by the electromagnetic coupling constant. 
Since the temporal component of the vector field $A_{0}$ is negligibly small, we consider only its spatial components $\vec{A} = (A_x, A_y, A_z)$ in the following discussion.

Assuming the standard halo model, the local density of DM is $\rho_{\rm DM} \sim 0.4\, {\rm GeV/cm}^3$, and its virial velocity is $v_{\rm vir}\simeq 220~{\rm km/sec}$ around the solar system in our Galaxy~\cite{Bertone:2004pz,Evans:2018bqy}. These profiles imply that DM would have an extremely large number density for the mass range $10^{-22} {\rm eV}/c^2 \lesssim m \lesssim 1 {\rm eV}/c^2$ and behave as a classical wave oscillating at about the Compton frequency $f_c = m_Ac^2/2\pi\hbar$. 
Then, one can describe it as the superposition of plane waves with different velocities $\vec{v}_{(i,n)}$ and phases $\theta_{(i,n)}$ as
\begin{equation}
\begin{aligned}
A_i(t, \vec{x}) &= \frac{A}{\sqrt{N}} \sum_{n = 1}^N \cos \left(2 \pi f_c \left(1 + v^2_{(i, n)}/2c^2\right) t \right. \\
 &- \left. \hbar^{-1}m_A\vec{v}_{(i,n)}^{} \cdot \vec{x} + \theta_{(i, n)}\right),\label{eq_DMwave}
\end{aligned}
\end{equation}
where $\theta_{(i,n)}$ is a random variable following a uniform distribution over $[0,2\pi]$ and $A \equiv \sqrt{2\rho_{\rm DM}\hbar^2/\varepsilon_0 3m_A^2c^4}$. 
Note that, for simplicity, the absence of correlation between the direction of $\vec{A}$ and $\vec{v}$ is assumed. 
Here $\vec{v}_{(i,n)}$ follows the DM velocity distribution of the standard halo model~\cite{Evans:2018bqy}:
\begin{equation}
\begin{aligned}
	f_{\rm SHM}(\vec v) ~\df^3 \vec{v}
	&= \frac{1}{(\pi v_{\rm vir}^2)^{3/2}}
	\exp\left[- \frac{(\vec{v} + \vec{v}_{\odot})^2}{ v_{\rm vir}^2}\right]
    ~\df^3 \vec{v},
    \label{eq_SHM_vec}
\end{aligned}
\end{equation}
with the solar velocity $|\vec{v}_{\odot}| \simeq 232\, {\rm km/sec}$. 
This distribution results in the velocity dispersion of the DM as $\bar{v}^2 = v_{\odot} ^2 + (3/2)v_{\rm vir}^2 \simeq \mathcal{O}(10^{-6}c^2)$. 
In such a non-relativistic regime, the time derivative of the field dominates over the spatial derivative. Hence, it can be regarded as a oscillating dark ``electric'' field, which induces displacements of 
the test masses in GW interferometers as $\delta\ddot{x}_i = \epsilon_De(Q/M)\dot{A}_i$. 

Another feature is that there appears a length scale, called ``coherence length'', which is evaluated as $L = 2\pi\hbar/m_A\sqrt{\bar{v}^2} \sim 10^{7}$km $(10^{-13}{\rm eV}\cdot c^{-2}/m_A)$.
This scale characterizes the spatial variation of the ultralight DMs.
As we will see below, the separation of test masses here is much shorter than $L$, we can take $\vec{x} = 0$ in Eq.~\eqref{eq_DMwave} without loss of generality and hereafter neglect the position dependence of the vector field.

\subsection{Auxiliary length channels of KAGRA}
\begin{figure}[htbp]
  \centering
  \includegraphics[width=\columnwidth]{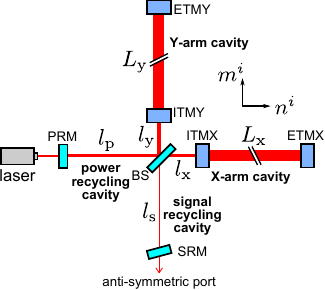}
  \caption{The schematic of the KAGRA interferometer. ITM (ETM): input (end) test mass, BS: beam splitter, PRM: power recycling mirror, SRM: signal recycling mirror.}
  \label{fig:IFO}
\end{figure}

From 2020 April 7 to 2020 April 21, KAGRA conducted its first joint observation run (O3GK)~\cite{KAGRA:2022twx, KAGRA:2022fgc}, together with GEO600.
The configuration of KAGRA interferometer during O3GK is shown in Fig.~\ref{fig:IFO}.
Similarly to that of LIGO and Virgo, it is based on a Michelson interferometer with a Fabry-P{\'e}rot cavity in two perpendicular arms. Each arm cavity is formed by the input test mass (ITM) and the end test mass (ETM). 
The main channel to monitor the differential changes caused by GWs is called the differential arm length (DARM).
The differential length between the beam splitter (BS) and two ITMs are controlled so that the Michelson fringe will be at the dark fringe at the anti-symmetric port, and the channel to monitor the differential Michelson interferometer length is called MICH. The power recycling mirror (PRM) and two ITMs form a power recycling cavity to effectively enhance the input power. The channel to monitor the power recycling cavity length is called PRCL.

During O3GK, a signal recycling mirror (SRM) was tilted, and the signal recycling cavity was not formed~\cite{KAGRAPTEP01}. Instead, this tilted SRM introduced an optical loss of 70\%, which led to degraded shot noise for the DARM readout. Two auxiliary channels, MICH and PRCL were recorded using different interferometer sensing ports, and were not affected by this optical loss. Using the length symbols in Fig.~\ref{fig:IFO}, changes in DARM, MICH and PRCL can be written as
\begin{eqnarray}
 \delta L_{\rm DARM} &=& \delta  (L_{\rm x} - L_{\rm y}), \\ \label{eq:darm}
 \delta L_{\rm MICH} &=& \delta (l_{\rm x} - l_{\rm y}), \\ \label{eq:mich}
 \delta L_{\rm PRCL} &=& \delta [(l_{\rm x} + l_{\rm y})/2 + l_{\rm p}],  \label{eq:prcl}
\end{eqnarray}
respectively. Note that each length parameter is given as $L_x = L_y = 3000$m, $l_x = 26.7$m, $l_y = 23.3$m and $l_p = 41.6$m, all of which are much shorter than the coherence length of DM.

Previous vector DM searches using LIGO and Virgo focused on DARM channels with the highest displacement sensitivity. 
This sensitivity is equivalent to the sensitivity to vector DM interactions for both LIGO and Virgo employing room temperature fused silica mirrors for all the mirrors.
This situation, however, drastically changes for KAGRA, which employs cryogenic sapphire mirrors for the test masses. 
As pointed out in Ref.~\cite{Michimura:2020vxn}, MICH and PRCL contain both the fused silica mirrors such as PRM and BS, and the sapphire test masses that respond differently to the vector DM due to the different charge-to-mass ratio.  
This results in the enhancement of the (differential) length change caused by vector DM in those channels.
Since the difference becomes larger especially for the $U(1)_{B-L}$ gauge boson, hereafter we focus on the $D = B - L$ case.

\subsection{Signal in the KAGRA's auxiliary length channels}\label{signal}
Since the frequency of each plane wave is localized near the Compton frequency as $f_i = f_c\sqrt{1+v_i^2/c^2} \sim f_c(1 + \mathcal{O}(10^{-6}))$, it is convenient to work in the Fourier space. 
Let us consider the Fourier transform of the signal in $X$ channel with duration $T$ and center time $t_0$:
\begin{equation}
    \tilde{h}_{X} (f; t_0) \equiv \int_{t_0 - T / 2}^{t_0 + T / 2} dt\> h_X(t) \e^{-2 \pi \iu f (t - t_0 + T / 2)}.
\end{equation}
To simplify the discussion, $T$ is taken to be much shorter than the time scale of Earth's rotation. 
As discussed in~\cite{Nakatsuka:2022gaf}, the oscillating length changes of an arm cavity can be decomposed into three contributions referred to as charge asymmetry, spatial difference and finite light traveling time~\cite{Morisaki:2020gui}. For auxiliary length monitors of interest here, the dominant contribution is the one from charge asymmetry expressed as 
\begin{equation}
\tilde{h}_{X} (f; t_0) = \iu \frac{\epsilon_D e}{2 \pi f} \Delta \left(\frac{Q_D}{M}\right) l^i_X(t_0) \tilde{A}_i(f; t_0), \label{eq:hX}
\end{equation}
where $\Delta(Q_D/M)$ is the difference of charge to mass ratio between two mirrors (BS and ITMX/Y) and in our case, it is $\Delta(Q_{B-L}/M) \sim 0.009 / m_n$ with $m_n$ being the neutron mass~\cite{Michimura:2020vxn}.
The vector $l^i_X(t_0)$ is given as
\begin{equation}
l^i_X(t) = \begin{cases}
\displaystyle n^i (t) - m^i (t), & (X=\text{MICH}) \\
\displaystyle \frac{1}{2} \left(n^i (t) + m^i (t)\right), & (X=\text{PRCL}),
\end{cases}
\end{equation}
where $n^i$ and $m^i$ are unit vectors pointing along the orthogonal arm axes depicted in Fig.~\ref{fig:IFO}.
Here $\tilde{A}_i(f; t_0)$ is the Fourier transform of the field amplitude. Since it is a superposition of a huge number of partial waves, the central limit theorem assures that $\tilde{A}_i(f; t_0)$ and consequently $\tilde{h}_{X} (f; t_0)$ follow a Gaussian distribution.
Note that, when the DM density and the distance between the test masses are fixed, this type of contribution results in a larger field amplitude and hence in a larger signal for the lower frequencies~\cite{Michimura:2020vxn,Nakatsuka:2022gaf}.
This is the reason why MICH and SRCL (not available in O3GK) has, under the design sensitivity, the capability of limiting the DM coupling beyond existing limits in the lower frequency band.

At this point, it is convenient to introduce the so-called coherence time
\begin{equation}
    \tau \equiv 2\pi/m_Av^2_{\rm vir} \sim 0.3~\mathrm{day}~\left(\frac{10^{-13}~{\rm eV}\cdot c^{-2}}{m_A}\right),
\end{equation}
which quantifies the characteristic correlation lifetime of the DM field at different times $\langle \tilde{h}_X^\ast(f;t_0) \tilde{h}_X(f;t_1)\rangle$. 
Let $T$ be the duration of a single chunk of data and $N_{\rm ch}$ be the number of equal-length chunks.
For higher mass ranges, in general, the coherence time becomes shorter than the duration of each chunk as $\tau < T$. 
In this case, the amplitude and phase of the DM randomly evolve within the chunks. 
While this decoherence reduces the growth of the signal-to-noise (SNR) ratio in amplitude as $\propto (N_{\rm ch}T)^{1/4}$~\cite{Nakatsuka:2022gaf}, each data chunk can be regarded as an independent measurement of DM, and in fact the statistical treatment can be simplified~\cite{Nakatsuka:2022gaf}. 
For a lower mass range ($\tau > T$), however, the correlation of the DM field between different data chunks cannot be neglected. This issue will be addressed in the following section describing the search pipeline.

Finally, let us give a concrete expression of the signal covariance $\langle \tilde{h}_X^\ast(f;t_0) \tilde{h}_X(f;t_1)\rangle$ used in our analysis.
By combining Eqs.~\eqref{eq_DMwave} and ~\eqref{eq_SHM_vec}, it is derived as
\begin{equation}
\begin{aligned}
&\left< \tilde{h}_X^\ast(f;t_0) \tilde{h}_X(f;t_1)\right> = \frac{\epsilon_D^2 e^2 A^2 T^2 \vvir^3}{32 \pi^{\frac{5}{2}} f^2 v_{\odot}^3} \left\{ \Delta \left(\frac{Q_D}{M}\right) \right\}^2\times\\ &l^i_X(t_0) l_{X, i}(t_1)\e^{-\frac{v_{\odot}^2}{\vvir^2} + 2 \pi \iu \fDM (t_1 - t_0)} \left(I(x_+) - I(x_-)\right).\label{eq:DM_cov}
\end{aligned}
\end{equation}
Here $I(x)$ is a function of the DM frequency and the time
\begin{equation}
\begin{aligned}
I(x) &= \frac{X^2}{8} \left[\sqrt{\pi} X \e^{X^2 / 4} \left\{ \erf \left(\frac{x}{X} - \frac{X}{2}\right)\right.\right.\\
&+ \left.\left. \erf \left(\frac{x}{X} + \frac{X}{2} \right) \right\}  -4 \e^{- x^2 / X^2} \sinh(x) \right],
\end{aligned}
\end{equation}
where we use the following parametrization,
\begin{equation}
x = \frac{2 v_{\odot}}{\vvir^2} v,~~~X = \frac{2 v_{\odot}}{\vvir \sqrt{1 - \iu \pi \vvir^2c^{-2} \fDM (t_1 - t_0)}},
\end{equation}
and
\begin{equation}
\frac{v_{\pm}}{c} \equiv \sqrt{2 \left(\frac{f \pm 1/(2T)}{\fDM} - 1\right)},~~~~~~x_{\pm} \equiv \frac{2 v_{\odot}}{\vvir^2} v_{\pm}.
\end{equation}
For $t_0 = t_1$, our $I(x)$ coincides with the function $\Delta_s(f_{n})$ in Ref.~\cite{Nakatsuka:2022gaf}, which gives the deterministic part of the spectral shape. 

\section{Search Method}\label{method}

\subsection{Detection statistics}
Here we introduce, and slightly extend, the detection statistic discussed in Ref.~\cite{Nakatsuka:2022gaf}. It is based on the existing methods of continuous GW searches~\cite{PhysRevD.57.2101, PhysRevD.58.063001, PhysRevD.61.082001} that also look for a narrow band signal similarly to our case. For comprehensive reviews of those method, see {\it e.g.} Refs.~\cite{Tenorio:2021wmz, Riles:2022wwz, Wette:2023dom}
The search method discussed in Ref.~\cite{Nakatsuka:2022gaf} is generally applicable to the ultralight DM searches using a single detector and interested readers could refer to Ref.~\cite{Oshima:2023csb} for the application of this detection statistic to the ultralight axion search.

As we discussed in Sec.~\ref{model}, the spectrum of the DM signal is localized within the narrow frequency band $f_c \leq f \lesssim f_c(1 +\kappa^2v^2_{\rm vir}/c^2)$ where $\kappa$ is $\mathcal{O}(1)$ constant and we set $\kappa=3.17$ to guarantee 
that the fractional loss of signal power becomes 
less than $1\%$.
For each single data chunk, we sum up the spectra over this frequency range: 
\begin{equation}
\rho_i(f_c) \equiv \sum_{f_c \leq f_n \leq f_c(1 +\kappa^2v^2_{\rm vir}/c^2)}\frac{4|\tilde{d}(f_n ; t_i)|^2}{TS(f_n; t_i)},
\end{equation}
where $T$ again is the duration of the data chunks, $\tilde{d}(f_n; t_i)$ represents the Fourier transform of the $i$-th data chunk and $S(f_n; t_i)$ is the one-sided noise power spectral density (PSD) around $t = t_i$.
Note that in order to neglect the effect of Earth's rotation as in Sec.~\ref{model}, $T$ should be small enough and hereafter we take $T = 30$ min.

In our pipeline, $S(f_n; t_i)$ is estimated from $\tilde{d}(f_n; t_i)$ by applying the running median and then converting it to the mean value by multiplying a correction factor (see App.A of Ref.~\cite{LIGOScientific:2005rkg}). In the median estimation, 180 neighboring frequency bins corresponding to $0.1$Hz band width are involved so that the effect of DM signal with narrow bandwidth can be smeared out. The number of bins involved in $\rho_i(f_c)$ is given as
\begin{equation}
    N_\mathrm{bin} 
    = 
      \left\lceil
    \frac{ \kappa \bar v^2f_c}{ \Delta f}
    \right\rceil
    =
    \left\lceil
    \kappa 
    \frac{T}{\tau }
    \right\rceil
    ,
    \label{eq_num_bin}
\end{equation}
where $\lceil x\rceil$ represents the minimum integer larger than $x$. 
By performing the summation over all chunks, we can define the detection statistics $\rho$ as
\begin{equation}
    \rho(f_c) \equiv \sum_i^{N_{\rm ch}}\rho_i(f_c),\label{eq:rho_tot}
\end{equation}
where $N_{\rm ch}$ represents the number of chunks.
Under the assumption of the stationarity and Gaussian distribution of noise, $\rho$ follows a $\chi^2$ distribution with $2N_\mathrm{bin}N_\mathrm{ch}$ degrees of freedom in the absence of signal. We chose the threshold to be the $95\%$ percentile of this distribution.

\subsection{Upper limit estimation}
In this study, we derive upper limits based on the frequentist's method where $\beta$ \% confidence level upper limit is derived through the integration of the likelihood function $\mathcal{L}(\rho(f_c);\epsilon_D^{\beta \%})$ as
\begin{equation}
    1 - \frac{\beta}{100} = \int^{\rho_{obs}}_0 d\rho\mathcal{L}(\rho(f_c);\epsilon_D^{\beta \%}).
\end{equation}
One might expect the central limit theorem to be applicable to $\rho(f_c)$ since the number of chunks is relatively large. There is, however, a non-vanishing cross-correlation between different segments $\left.\langle \rho_i\rho_j \rangle\right|_{i\neq j} \neq 0$ for $|t_i - t_j| < \tau$. This correlation prevents the convergence to a Gaussian distribution especially for lower-mass DM and makes the analytical expression of $\mathcal{L}(\rho(f_c);\epsilon_D)$ complicated~\cite{Nakatsuka:2022gaf}. In fact, this analytical expression suffers from numerical instability for intermediate regimes where the coherence time and the duration of chunks are comparable.
In our pipeline, therefore, the 95\% upper limit on the coupling constant was numerically derived as follows.

Assuming that only Gaussian noise $\tilde{n}(f_n ; t_i)$ and ultralight DM signals are present in the data, it can be expressed as $\tilde{d}(f_n ; t_i) = \tilde{n}(f_n ; t_i) + \tilde{h}_X(f_n ; t_i)$. Therefore, 
dependence of $\rho$ on the coupling constant can be decomposed as
\begin{equation}
\begin{aligned}
    \rho(f_c; \epsilon_D) 
    &= \sum_{t_i, f_n}\frac{4}{TS}\left(|\tilde{n}|^2 + 2{\rm Re}\left[\tilde{n}^*\tilde{h}_X\right] + |\tilde{h}_X|^2\right)\\
    &= \mathcal{N}^2 + \epsilon_D \mathcal{N}\cdot\mathcal{S} + \epsilon_D^2 \mathcal{S}^2.
\end{aligned}
\end{equation}
Here $\mathcal{N}^2$ represents contributions from $2N_{\rm bin}N_{\rm ch}$ unit Gaussian variables since the noise component in $\rho$ is normalized by the PSD.
On the other hand, $\mathcal{S}^2$ represents contributions solely from the Gaussian signal $\tilde{h}_X(f_n ; t_i)$ (also normalized by noise PSD) whose correlation function is given as Eq.~\eqref{eq:DM_cov} under the standard halo model assumption.
$\mathcal{N}\cdot\mathcal{S}$ is the contribution from the cross term of the unit Gaussian and the normalized signal.
For a fixed DM mass (or $f_c$), we simulated $10^5$ realizations of $\mathcal{N}^2,\ \mathcal{S}^2,\ \mathcal{N}\cdot\mathcal{S}$ from the covariance of the DM signal~\eqref{eq:DM_cov} and the estimated noise PSD $S_n(f_n;t_i)$.
Then we can obtain a histogram of $\rho$ that approximates the likelihood $\mathcal{L}(\rho(f_c);\epsilon_D)$ and depends on the value of $\epsilon_D$.
Then the value of $\epsilon_D$, for which the observed value of the detection statistics $\rho_{\rm obs}(f_c)$ coincides with the 5\% percentile of this realization, is identified as the 95\% upper limit.
We would like to emphasize that our method does not suffer from the numerical instability mentioned above and that it is applicable to arbitrary masses of DM.

\section{Analysis}\label{analysis}
\subsection{Data}
We analyzed the O3GK data collected from the KAGRA detector in the observation mode, and the data from GEO600 is not used. During the O3GK run, the KAGRA detector had duty factors, the fraction of time the detector is in the observing mode to the total time, of $\sim 53\%$.
The length channels considered in this study are the differential Michelson interferometer length (MICH) and the power recycling cavity length (PRCL).
For both the MICH and PRCL channels, calibrations of data were performed offline for this study, whose parameters and information will be summarized in Ref.~\cite{opendata}. 
For the frequency band we used for the analysis, calibration uncertainty in MICH and PRCL channels are 20-30\% in amplitude.

\begin{figure}[htbp]
  \centering
  \includegraphics[width=\columnwidth]{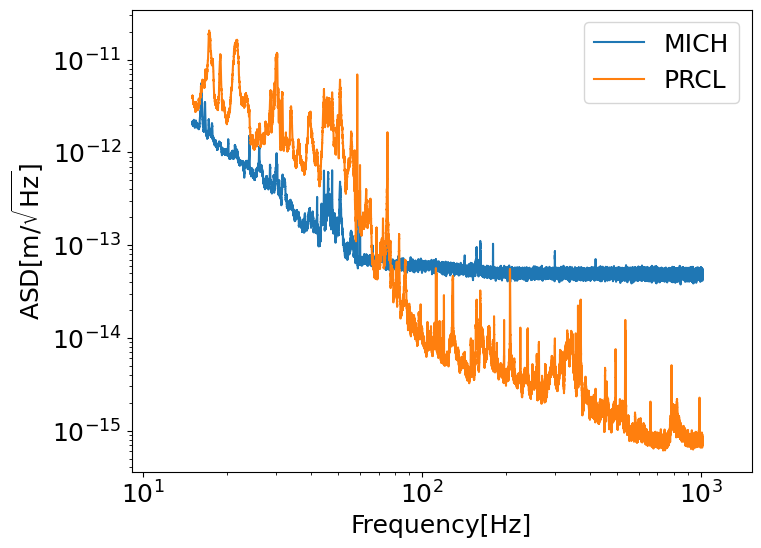}
  \caption{ASDs of the MICH and PRCL estimated from the first 30 min chunk during O3GK.}
  \label{fig:ASDs}
\end{figure}

The amplitude spectral densities (ASDs), derived from $S(f_n;t_i)$, of these channels during O3GK are plotted in Fig~\ref{fig:ASDs}.
We should note that, for the last few days of O3GK, alignments of mirrors were dithered at significantly large amplitude for the beam position control~\cite{KAGRA:2022fgc}. These injected lines were accompanied by a large number of sidebands. Since our pipeline simply searches power excesses within narrow bandwidth, it is not straightforward to distinguish those noise lines from the DM signals.
Hence the segments from the last three days of the O3GK, where the efficiency of the DM search was spoiled, were not included in our analysis. Consequently, the number of 30 minute chunks subject to our pipeline was 217. 

Another limitation is that the vibration isolation systems for mirrors in MICH and PRCL were simplified compared to those in the full-design~\cite{KAGRA:2022fgc}. Consequently, there are many noise peaks in the lower frequency range, where the KAGRA's auxiliary degrees of freedom become more and more sensitive. 
Although our pipeline can be used in the lower frequency range $\lesssim 10$Hz, the analysis was performed over the frequency range from 15~Hz to 1015~Hz for this demonstration analysis.

\subsection{Candidates and veto procedure}
\begin{figure}[tb]
  \centering
  \includegraphics[width=\columnwidth]{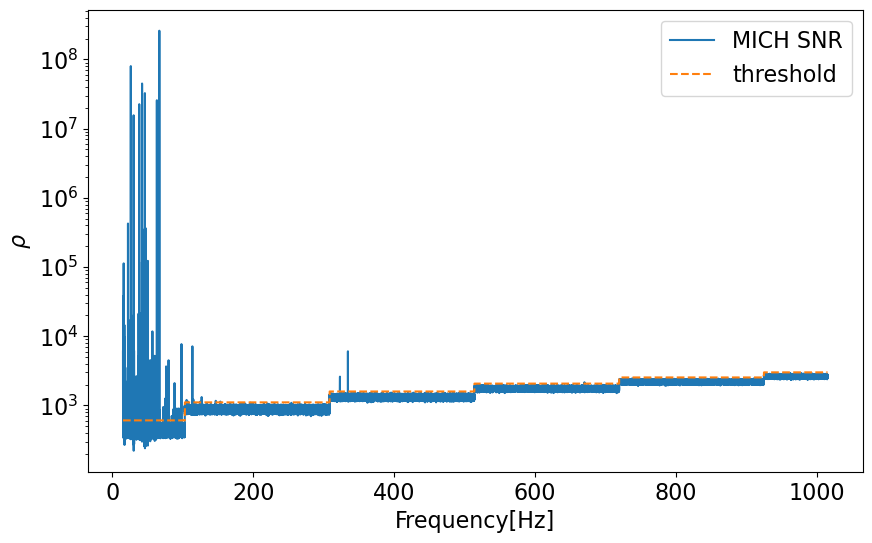}\\
  \includegraphics[width=\columnwidth]{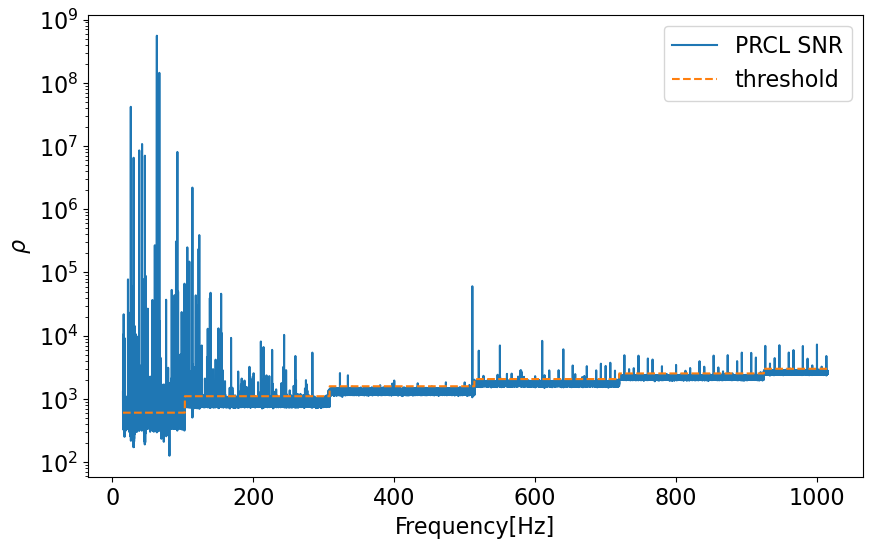}
  \caption{The detection statistics $\rho$ computed from 217 half-hour chunks during O3GK with MICH ({\it top panel}) and PRCL ({\it bottom panel}) channel. A step appears about every 200 Hz because of the increase in $N_{\rm bin}$, which varies from 1 to 6. The step height depends on the number of chunks. 
  }
  \label{fig:SNRs}
\end{figure}

\begin{figure}[tb]
  \centering
  \includegraphics[width=\columnwidth]{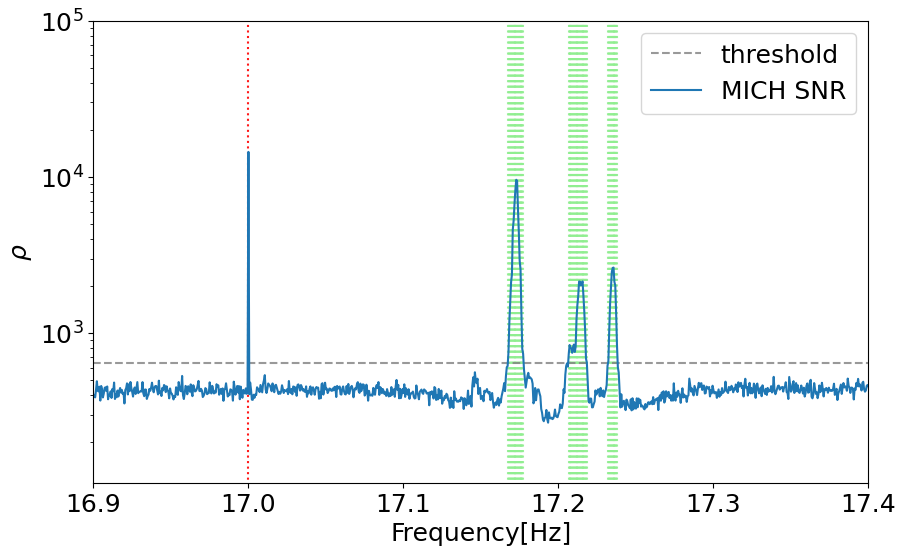}\\
  \includegraphics[width=\columnwidth]{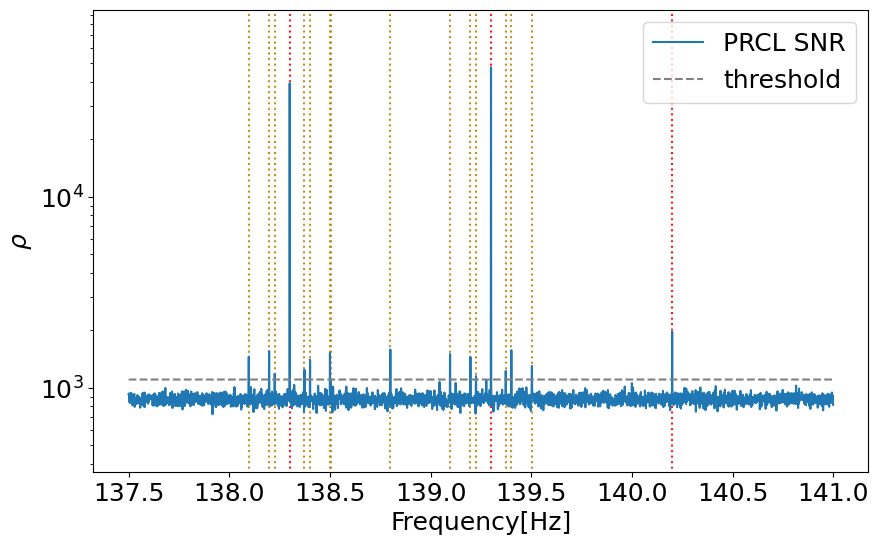}
  \caption{A schematic picture of the veto procedure. For both cases, the vertical red lines denote surviving candidates after two veto procedure. ({\it Top panel}) Green vertical bands indicate the broad peak structures vetoed by the bandwidth criterion. ({\it Bottom panel}) The yellow vertical lines represent candidates with power excess that fails coincidence requirements.}
  \label{fig:veto}
\end{figure}

In Fig.~\ref{fig:SNRs}, the detection statistics $\rho(f_c)$ computed in our pipeline are shown respectively for MICH and PRCL. As expected from the many lines in ASDs shown in Fig.~\ref{fig:ASDs}, more non-Gaussian power excess within an expected signal bandwidth is observed in the PRCL data.
Under the $5\%$ False-Alarm-Probability derived from $\chi^2$ distribution for Gaussian noise, 1944 and 4133 lines are identified as candidates for MICH and PRCL, respectively.
These candidates are then subject to the veto analysis as follows.
First, according to the expected narrow band feature of DM signal, the peak of $\rho(f_c)$ should also have a comparable width if it has DM origin.
As illustrated in the top panel of Fig.~\ref{fig:veto}, candidates with a width more than two times broader than the expected DM signal width were vetoed. 
We also vetoed lines that were less than the peak width away from the vetoed peaks of $\rho$, as there is a possibility that they originated from the same source as those broad peaks.

Second, in contrast to transient noise, expected to produce outliers in the O3GK data, DM signals should be more persistent on average, albeit with random statistical fluctuations.
In the analysis, we chronologically divide the whole data chunks into two subsets and perform the same analyses within the first half 108 chunks and the second half 108 chunks.
Then we take a coincidence between the candidates found in those two subsets to exclude transients that only appear during a limited duration.
However, it must be noted that this procedure inevitably vetoes weaker DM signals with which $\rho(f_c)$ exceeds the detection threshold only after summing over all the chunks.
Furthermore, unless the signal coherence time is sufficiently shorter than the data length of the subset, the field amplitude (and therefore the value of $\rho(f_c)$) can be significantly different for each subset.
By examining the spectrum, we found that in the case of present data, such lines with the values of $\rho$ close to the threshold are mostly sidebands of more intense lines, but this issue must be considered in future studies.
For possible improvements to this veto procedure, see the discussion below.

After applying these two veto procedures, 77 lines remain as candidates in the MICH data while 202 lines remain in the PRCL data.
Many of them are in the low frequency range below 100~Hz, where there are many lines due to injected signals and the suspension noise.
By referring to, for example, the Appendix A. of Ref~\cite{KAGRA:2022fgc} discussing the noise lines of DARM during O3GK, we found that 57 out of 77 lines in the MICH data and 54 out of 202 lines in the PRCL data come from known lines.
Note that there were few dedicated studies of the line identification for MICH and PRCL channels during O3GK, and lines of unknown origin still remain.
Therefore, for our future reference, the candidate lines identified in this study are listed in Ref.~\cite{vDM_lines}. 

\subsection{Upper limit}
There are several lines that have passed our pipeline detection criteria, but they cannot be claimed as DM signals given the current level of displacement sensitivity shown in Fig.~\ref{fig:ASDs} and the much shorter total observation time, compared to those assumed in Ref.~\cite{Michimura:2020vxn}. 
With the scaling of ${\rm SNR} \propto (N_{\rm ch}T)^{1/4}$, their prediction on upper limits derived by setting ${\rm SNR} = 1$ can be scaled to give an order of magnitude estimate of the upper limits we can derive from $\rho$. 
By comparing the design sensitivity (Fig.~2 in Ref.~\cite{Michimura:2020vxn}) with our Fig.~\ref{fig:ASDs} and considering another suppression factor of $(N_{\rm ch}\cdot30{\rm min}/1{\rm yr})^{1/4} \sim 0.01^{1/4}$ that comes from the difference between the real and assumed observation time, we expect that the upper limit on $\epsilon_{B - L}$ from O3GK data will be at best $\mathcal{O}(10^{-20})$. This is much weaker than those derived from the E\"{o}t-Wash torsion balance~\cite{Schlamminger:2007ht, Wagner:2012ui} and MICROSCOPE~\cite{Touboul:2017grn, Berge:2017ovy, Fayet:2017pdp}, predicting $10^{-24}$ to $10^{-23}$ in the frequency band of our interest.

Therefore, the upper limit estimation here should be considered as a demonstration of our pipeline, which can accurately analyze low-frequency regions and high-frequency regions as well.
As such a demonstration, upper limits on the coupling constant are derived over the whole frequency range analyzed in this study. Note that in principle, upper limits can be derived even for $\rho(f_c)$ that exceed the detection threshold.

\begin{figure}[htbp]
  \centering
  \includegraphics[width=\columnwidth]{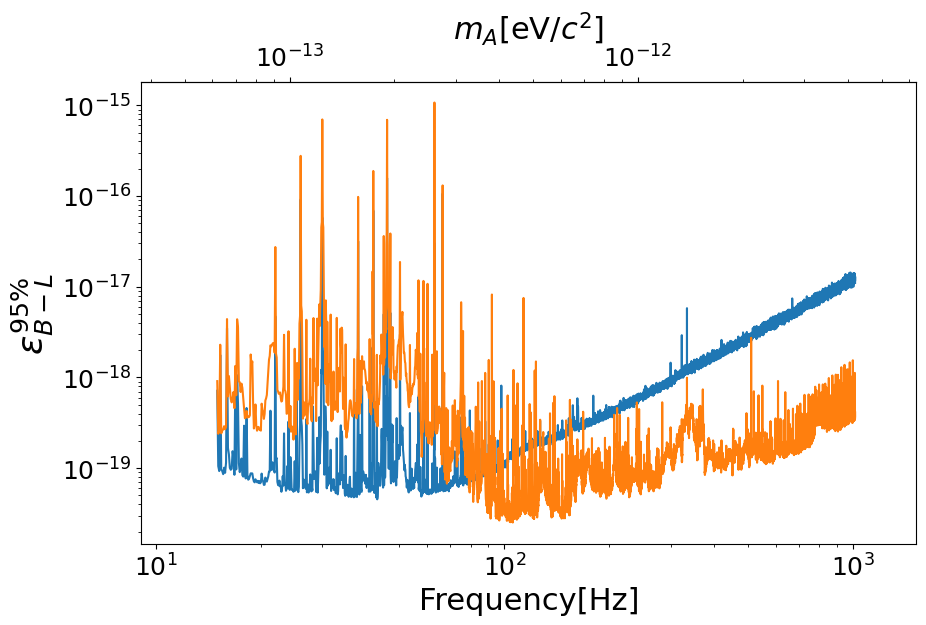}
  \caption{95\% upper limit on the $B-L$ gauge coupling constant derived from MICH data (blue line) and PRCL data (orange line). Many narrow peaks observed in lower mass range are due to unknown line artifacts in the lower frequency range.}
  \label{fig:UL}
\end{figure}

In Fig.~\ref{fig:UL}, 95\% upper limits on the coupling constant $\epsilon_{B-L}$ are shown. Here the constraint is smoothed by collecting the maximum value of $\rho(f_c)$ within a 0.1Hz bandwidth. 
For clarity, uncertainty of the calibration is not displayed. 
Let us stress that the stochastic nature of DM is properly taken into account in deriving this bound, by using the covariance of DM signal~\eqref{eq:DM_cov} for simulating the realization of $\rho(f_c)$. 
As demonstrated in Ref.~\cite{Nakatsuka:2022gaf}, for example, incorrect deterministic treatment predicts a non-central $\chi^2$ distribution of $\rho(f_c)$ and overestimates the upper bound by up to a factor of 3 in low mass (or long coherence time) regions.
We also found that our result is consistent with the above rough estimate on the upper limit based on the SNR scaling, which is also a consequence of the stochastic nature of ultralight DMs~\cite{Michimura:2020vxn, Nakatsuka:2022gaf}.

As expected, these upper limits derived from the KAGRA O3GK data 
are weaker than previous published limits by several orders of magnitude. 
Again, this is owing to both the current noise level of the KAGRA detector and the limited duration of the measurement time. 
In order to reach the unexplored parameter space, reduction of the dominant low-frequency noise and longer stable detector operation are indispensable. 

\section{Discussion}\label{discussion}
In this work, an ultralight vector DM search with KAGRA was conducted for the first time by using the KAGRA data from the O3GK run. Our pipeline design is guided by a recent study on the stochastic nature of the ultralight vector DM~\cite{Nakatsuka:2022gaf}. Consequently, the KAGRA O3GK data, which has a relatively short measurement time compared to the DM coherence time for low DM masses, can be analyzed.

We found that our pipeline can discriminate candidates for vector DM signal from the broad peaks and transient lines, which are expected to have an instrumental origin.
Nonetheless, there are several lines with unspecified origin meeting our criteria, especially for the lower frequency range.
We expect that the number of such lines will decrease in upcoming observations because of, for example, an updated system for the suspension control and an improved understanding of the noise. 
Although the upper limits derived in this analysis are weaker than previous ones, they are found to be consistent with the prediction given in prior studies~\cite{Michimura:2020vxn,Nakatsuka:2022gaf}.
Achieving the designed sensitivity and future upgrades of the KAGRA detector~\cite{KAGRAplus} will allow us to fully appreciate its unique feature as a vector DM detector exploiting the mirrors made of different materials to yield new constraints.

There are several directions for improving our DM search pipeline. First, as outlined in Sec.~\ref{model}, our analysis is performed assuming the equilibration of vector polarization.
Depending on the production mechanism, however, only a specific polarization mode might be produced.
Therefore, the formalism beyond this assumption may allow us to probe the cosmological origin of vector DM.
Second, while the bandwidth criterion for the veto analysis is quite robust, the appropriateness of taking coincidence between subsets of whole data should depend on the strength and coherence time of the signal being searched for.
Since the covariance of the vector DM signal is known, it is possible to implement a test to check whether the candidates follow a distribution consistent with the given covariance. This could provide a more robust way to distinguish a DM signal from the noise lines. 
Finally, our pipeline can be extended to include the analysis of the DARM channel data, which has been used in DM searches in other GW detectors.

\section*{Acknowledgements}
This material is based upon work supported by NSF’s LIGO Laboratory which is a major facility
fully funded by the National Science Foundation.
The authors also gratefully acknowledge the support of
the Science and Technology Facilities Council (STFC) of the
United Kingdom, the Max-Planck-Society (MPS), and the State of
Niedersachsen/Germany for support of the construction of Advanced LIGO 
and construction and operation of the GEO\,600 detector. 
Additional support for Advanced LIGO was provided by the Australian Research Council.
The authors gratefully acknowledge the Italian Istituto Nazionale di Fisica Nucleare (INFN),  
the French Centre National de la Recherche Scientifique (CNRS) and
the Netherlands Organization for Scientific Research (NWO), 
for the construction and operation of the Virgo detector
and the creation and support  of the EGO consortium. 
The authors also gratefully acknowledge research support from these agencies as well as by 
the Council of Scientific and Industrial Research of India, 
the Department of Science and Technology, India,
the Science \& Engineering Research Board (SERB), India,
the Ministry of Human Resource Development, India,
the Spanish Agencia Estatal de Investigaci\'on (AEI),
the Spanish Ministerio de Ciencia e Innovaci\'on and Ministerio de Universidades,
the Conselleria de Fons Europeus, Universitat i Cultura and the Direcci\'o General de Pol\'{\i}tica Universitaria i Recerca del Govern de les Illes Balears,
the Conselleria d'Innovaci\'o, Universitats, Ci\`encia i Societat Digital de la Generalitat Valenciana and
the CERCA Programme Generalitat de Catalunya, Spain,
the National Science Centre of Poland and the European Union – European Regional Development Fund; Foundation for Polish Science (FNP),
the Swiss National Science Foundation (SNSF),
the Russian Foundation for Basic Research, 
the Russian Science Foundation,
the European Commission,
the European Social Funds (ESF),
the European Regional Development Funds (ERDF),
the Royal Society, 
the Scottish Funding Council, 
the Scottish Universities Physics Alliance, 
the Hungarian Scientific Research Fund (OTKA),
the French Lyon Institute of Origins (LIO),
the Belgian Fonds de la Recherche Scientifique (FRS-FNRS), 
Actions de Recherche Concertées (ARC) and
Fonds Wetenschappelijk Onderzoek – Vlaanderen (FWO), Belgium,
the Paris \^{I}le-de-France Region, 
the National Research, Development and Innovation Office Hungary (NKFIH), 
the National Research Foundation of Korea,
the Natural Science and Engineering Research Council Canada,
Canadian Foundation for Innovation (CFI),
the Brazilian Ministry of Science, Technology, and Innovations,
the International Center for Theoretical Physics South American Institute for Fundamental Research (ICTP-SAIFR), 
the Research Grants Council of Hong Kong,
the National Natural Science Foundation of China (NSFC),
the Leverhulme Trust, 
the Research Corporation,
the National Science and Technology Council (NSTC), Taiwan,
the United States Department of Energy,
and
the Kavli Foundation.
The authors gratefully acknowledge the support of the NSF, STFC, INFN and CNRS for provision of computational resources.

This work was supported by MEXT, JSPS Leading-edge Research Infrastructure Program, JSPS Grant-in-Aid for Specially Promoted Research 26000005, JSPS Grant-inAid for Scientific Research on Innovative Areas 2905: JP17H06358, JP17H06361 and JP17H06364, JSPS Core-to-Core Program A. Advanced Research Networks, JSPS Grantin-Aid for Scientific Research (S) 17H06133 and 20H05639 , JSPS Grant-in-Aid for Transformative Research Areas (A) 20A203: JP20H05854, the joint research program of the Institute for Cosmic Ray Research, University of Tokyo, National Research Foundation (NRF), Computing Infrastructure Project of Global Science experimental Data hub Center (GSDC) at KISTI, Korea Astronomy and Space Science Institute (KASI), and Ministry of Science and ICT (MSIT) in Korea, Academia Sinica (AS), AS Grid Center (ASGC) and the National Science and Technology Council (NSTC) in Taiwan under grants including the Rising Star Program and Science Vanguard Research Program, Advanced Technology Center (ATC) of NAOJ, and Mechanical Engineering Center of KEK.

Additional acknowledgements for support of individual authors may be found in the following document: \\
\texttt{https://dcc.ligo.org/LIGO-M2300033/public}.
For the purpose of open access, the authors have applied a Creative Commons Attribution (CC BY)
license to any Author Accepted Manuscript version arising.
We request that citations to this article use 'A.G. Abac {\it et al.} (LIGO-Virgo-KAGRA Collaboration), ...' or similar phrasing, depending on journal convention.

\bibliography{O3GKvDM}
\bibliographystyle{apsrev4-1}
\end{document}